\documentclass[prl,twocolumn,aps,showpacs]{revtex4}
\usepackage{graphicx}

\begin{document}

\title{
Single electron-phonon interaction in a suspended quantum 
dot phonon cavity 
      }
\author{E.~M. H\"ohberger$^{1}$}
\email{hoehberger@lmu.de}
\author{R.~H. Blick$^{1}$}
\altaffiliation{New present address: Electrical and Computer
  Engineering, University of Wisconsin-Madison, Madison, Wisconsin 53706, USA.}
\author{T. Brandes$^2$} 
\author{J. Kirschbaum$^1$} 
\author{W. Wegscheider$^3$}
\author{M. Bichler$^4$}
\author{J.~P. Kotthaus$^1$}
\affiliation{$^1$ Center for NanoScience \& Sektion Physik,
  Ludwig-Maximilians-Universit\"at,  
  80539 M\"unchen, Germany}
\affiliation{$^2$ Department of Physics, University of Manchester,
  Institute of Science and Technology (UMIST), Manchester M60 1QD,UK}
\affiliation{$^3$ Institut f\"ur Angewandte und Experimentelle Physik,
  Universit\"at Regensburg, 93040 Regensburg, Germany}
\affiliation{$^4$ Walter-Schottky-Institut, Technische Universit\"at
  M\"unchen, 85748 Garching, Germany}
\date{\today{ }}
\begin{abstract} 
An electron-phonon cavity consisting of a quantum dot embedded in a
free-standing GaAs/AlGaAs membrane is characterized in Coulomb blockade 
measurements at low temperatures.  We find a complete suppression of single 
electron tunneling around zero bias leading to the formation of an
energy gap in the transport spectrum. The observed effect is induced
by the excitation of a localized phonon mode confined in the
cavity. This phonon blockade of transport is lifted at magnetic fields
where higher electronic states with nonzero angular momentum are
brought into resonance with the phonon energy. 
\end{abstract}
\pacs{73.23.Hk, 73.63.Kv, 73.21.La, 71.38.-k, 62.25.+g, 85.85.+j}
   %73.21.La Quantum dots:
   %         -> Electron states and collective excitations in quantum dots
   %73.23.Hk Electronic transport in mesoscopic systems
   %         -> Coulomb blockade; single-electron tunneling
   %73.63.Kv Electronic transport in nanoscale materials and structures
   %         -> Quantum dots
   %71.38.-k polarons and electron-phonon interaction)             
   %62.25.+g mechanical properties of nanoscale materials
   %85.85.+j Micro- and nano-electromechanical systems (MEMS/NEMS) \& devices 
\maketitle

When Rolf Landauer discussed the importance of irreversibility and
heat generation for classical transistors in
1961~\cite{bib:landauer61}, he found that the minimal amount of
dissipation required to perform a single bit operation is given by
$k_{\rm B} T \ln 2$. In the context of quantum physics the role of dissipation
needs to be reconsidered. Even more, along with the discovery of
single electron transistors~\cite{bib:fulton87,bib:grabert92} and
quantum dots~\cite{bib:ashoori96,bib:kouwenhoven97}, it is nowadays
regarded as crucial for the feasibility of quantum
computation~\cite{bib:averin99,bib:bennett00}. The next step towards
control of dephasing of electronic quantum states is tailoring phonon
confinement in quantum dot cavities. The most promising approach to
address the physics of dissipation in the ultimate limit of single
electrons interacting with individual phonon modes of their host
crystal is to embed a low-dimensional electron gas into a suspended
phonon cavity as demonstrated in previous
work~\cite{bib:blick00,bib:hoehberger02,bib:kirschbaum02}. The
relevance of the electron-phonon interaction for quantum dot systems
was initially explored by Fujisawa et al.~\cite{bib:fujisawa98} and
Qin et al.~\cite{bib:qin01} and theoretically confirmed by Brandes and
Kramer~\cite{bib:brandes99}.

Already back in 1967, Duke et al.~\cite{bib:duke67} modelled inelastic
tunneling through a barrier, finding that collective phonon modes can
be excited by the tunneling electron. According to their calculations
characteristic zero bias conductance minima in the tunnel conductance
can be attributed to this effect. In recent years, electron
back-action on mechanical degrees of freedom has been theoretically
discussed by Schwabe et al.~\cite{bib:schwabe95} and
Blencowe~\cite{bib:blencowe99}. The additional implications of phonon
confinement in an electron-phonon cavity were modelled by Debald et
al.~\cite{bib:debald02}, pointing out the possibility to control
electron dephasing by tailoring the phonon spectrum.

Here we report on the experimental observation of a new blocking
mechanism of single electron transport which is found in such a
cavity, evidencing the coherent interplay between single electron
tunneling and the excitation of localized phonon modes confined in the
cavity as predicted by Duke et al.~\cite{bib:duke67}. Strikingly
similar features which can be attributed to the same effect have
been observed for single electrons tunneling onto a C$_{\rm
60}$ molecule by Park et al.~\cite{bib:park00,bib:mceuen02}. The
underlying physics of coherently coupling discrete electronic states
with discrete phonon modes bears resemblance to cavity
QED~\cite{bib:imamoglu99}.

\begin{figure}[t]
\centering
\includegraphics[width=7.5cm]{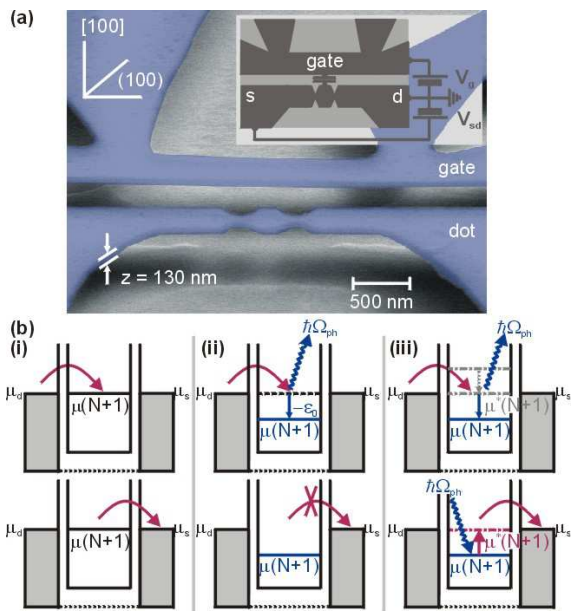}
\caption{
(a) Suspended quantum dot cavity and Hall-bar formed
in the blue-colored $130$\,nm thin GaAs/AlGaAs membrane.
The inset shows a schematic top view of the sample. 
(b) Level diagrams for single electron tunneling including 
phonon blockade: (i) In the orthodox model electrons sequentially
tunnel through the dot, if the  
chemical potential $\mu(N+1)$ is aligned between the reservoirs. (ii)
Tunneling into the phonon  
cavity  results in  the excitation of a cavity phonon with energy
$\hbar \Omega_{\rm ph}$, leading to a level 
mismatch $\epsilon_0$ and thus to phonon blockade. (iii) Single
electron tunneling
is re-established by a higher lying electronic state $\mu^*(N+1)$ 
which is enabled to coherently reabsorb the phonon and to hereby replace the
ground state~\protect\cite{footnote:ang_momentum}. 
} \label{fig1}
\vspace{-0.3cm}
\end{figure}

A phonon cavity containing a suspended single quantum dot is shown in
Fig. \ref{fig1}(a). The scanning electron micrograph was taken under
an angle of $65^\circ$ in order to visualize the three-dimensional
character of the sample. Depicted in blue is the free-standing
$130$\,nm thick GaAs/AlGaAs membrane containing a confined electron
gas which is located $40$\,nm below the sample surface. The $400$\,nm
thick sacrificial layer of Al$_{0.8}$Ga$_{0.2}$As supporting the
membrane has been completely removed beneath the displayed part of the
sample in Fig. \ref{fig1}(a) creating a spacing between the membrane
and the GaAs buffer displayed in grey. The quantum dot is defined in a
$600$\,nm wide bar by two point contacts formed by pairs of symmetric
indentations. As a result of edge depletion the cavity has a reduced
electronic diameter of about $450$\,nm~\cite{bib:kirschbaum02}. The
two constrictions are wide enough to allow ballistic transport through
the cavity, but can be depleted to form tunneling barriers by a nearby
gate electrode. In the case of Fig. \ref{fig1}(a) this task is
accomplished by a close-by Hall-bar being employed as an in-plane
gate. A schematic top-view of the sample layout is displayed in the
inset.

\begin{figure}[t]
\centering
\includegraphics[width=7.5cm]{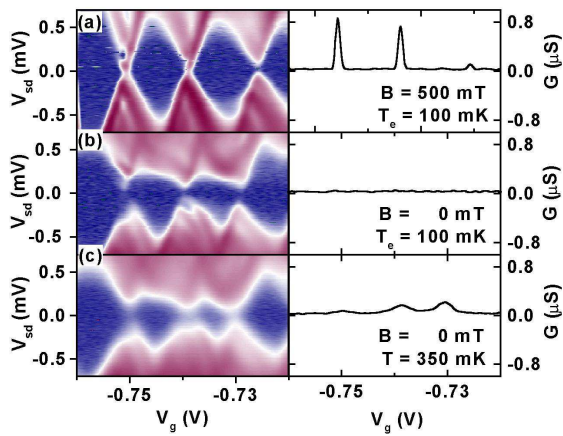}
\caption{
Transport spectrum of the suspended single quantum dot 
and zero bias conductance: (a) Single electron
resonances taken at an electron temperature of $100$\,mK and a 
perpendicular magnetic field of $500$\,mT. 
(b) At zero magnetic field conductance is suppressed for bias
voltages below $100\,\mu$V due to phonon excitation.
(c) The conductance pattern at $350$\,mK shows that phonon
blockade starts to be surpassed because of thermal broadening of the
Fermi function supplying empty states in the reservoirs.
}
\label{fig2}
\vspace{-0.3cm}
\end{figure}

\begin{figure}[t]
\centering
%\vspace{0.02cm}
\includegraphics[width=7.5cm]{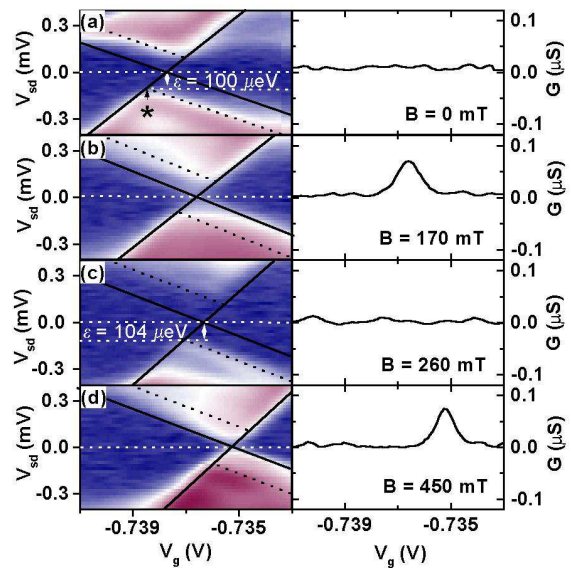}
\caption{
Transport spectrum for (a) $B = 0$\,mT, (b) $170$\,mT, (c) $260$\,mT,
and (d) $450$\,mT. 
The line plots
give the zero bias trace. At certain magnetic fields (b,d)
excited quantum dot states with higher magnetic momentum are brought
into resonance with the cavity phonon re-enabling single electron
tunneling. Otherwise (a,c) transport is suppressed due to phonon
blockade with an excitation barrier of around $100\,\mu$eV.
}
\label{fig3}
\vspace{-0.3cm}
\end{figure}

The presented measurements are performed in a dilution refrigerator
with a base temperature of $T_{\rm bath} = 10$\,mK. A negative voltage
$V_{\rm g}$ is applied to the gate electrode (i.e. the Hall-bar in
Fig. \ref{fig1}(a)) in order to create tunneling barriers and to vary
the electrochemical potential of the dot denoted \mbox{$\mu(N+1) = E(N+1) -
E(N)$ in} the level diagrams of Fig. \ref{fig1}(b) which will be
discussed in more detail below. A bias voltage $V_{\rm sd}$ can be
applied between the source and drain reservoirs. The differential
conductance $G = dI_{\rm sd}/dV_{\rm sd}$ is recorded with respect to
both $V_{\rm g}$ and $V_{\rm sd}$ showing clear Coulomb
diamonds~\cite{bib:kouwenhoven97} depicted in logarithmic color scale
representation in the left part of Fig. \ref{fig2} (blue: $0.02\,\mu$S, red:
$6\,\mu$S). The right part of the figure displays the corresponding
line plots at zero bias $V_{\rm sd} = 0\,\mu$V. Figure \ref{fig2}(a) was
taken at an electron temperature of $T_{\rm e} = 100$\,mK and a perpendicular
magnetic field of $B = 500$\,mT where a quasi-continuum of
energetically higher states produces a Coulomb blockade diamond known
from conventional quantum dots~\cite{bib:kouwenhoven97}. The charging
energy $E_{\rm C} = e^2/2C_\Sigma = 0.56$\,meV corresponds to a dot
capacitance of $C_\Sigma = 140$\,aF from which a dot radius of $r =
160$\,nm and an electron number of about $480$ can be deduced. The
striking difference as compared to conventional quantum dot
measurements is observed for the same temperature but zero magnetic
field where we find {\it complete} suppression of conductance around zero
bias~\cite{footnote:spinblockade}. This is shown exemplarily for three
adjacent resonances in Fig. \ref{fig2}(b), both from the gap opening
between the diamonds in the color scale plot and from the lineplot
showing zero conductance~\cite{footnote:diamondshape}. The resulting
blockade of transport can only be overcome by applying a positive or
negative source drain bias of $|V_{\rm sd}| = 100\,\mu$V (see also
Fig. \ref{fig3}(a) magnifying a single resonance).

These results can be compared to recent theoretical models for
transport through a molecular single electron transistor coupled to a
single vibrational mode~\cite{bib:boese01,bib:zhu02,bib:flensberg03}. We
have extended these models~\cite{bib:brandes03} in order to analyze
the origin of the observed blockade mechanism. The picture
corresponding to the classical limit (strongly overdamped vibration
mode) is illustrated in Fig. \ref{fig1}(b): (i) In case of a
conventional single electron transistor an electron sequentially
tunnels through a non-suspended quantum dot whenever Coulomb blockade
can be overcome, i.e. the electrochemical potentials of source, drain,
and the dot $\mu_s$, $\mu_d$ and $\mu(N+1)$ are aligned. This
situation corresponds to the well-known
M\"ossbauer-effect~\cite{bib:moessbauer64} in gamma spectroscopy:
Recoil-free absorption and emission of gamma-ray photons is enabled
for a nucleus being placed in a solid since the crystal takes up the
recoil as a whole without entailing an energy loss. In a classical
picture this is equivalent to the reflection of a particle at a hard
wall with infinite mass. (ii) This behavior changes dramatically for a
quantum dot embedded in a suspended phonon cavity a classical analogue
of which is given by a particle hitting a trampoline. Due to strong
electron-phonon coupling in the cavity~\cite{bib:debald02} (see
below), single electron tunneling induces mechanical displacement of
the suspended quantum dot which corresponds to the excitation of a
localized cavity phonon of energy $\hbar \Omega_{\rm ph}$. Flensberg has
shown~\cite{bib:flensberg03} that in the strongly overdamped,
classical limit (quality factor $Q=\Omega_{\rm ph}/\gamma \ll 1$ for small
phonon life time $\gamma^{-1}$), the energy cost of the displacement
goes along with a drop of the chemical potential $\mu(N+1)$ in the
dot, which leads to a blockade of single electron tunneling. The
energy gap $\epsilon_0 = g \hbar \Omega_{\rm ph}$ then depends on the
Franck-Condon coupling constant $g$. In the `M\"ossbauer
picture'~\cite{bib:duke67}, the cavity picks up the `recoil' of the
tunneling electron and immediately relaxes to a new ground state for
$Q \to 0$. On the other hand, $Q \to \infty$ would correspond to
coherent cavity phonons where the Franck-Condon factors yield a series
of phonon side-bands $n \Omega$ with weights given by the Poisson
distribution $e^{\rm -g} g^{\rm n}/n!$ at zero temperature with $n=0$
corresponding to elastic tunneling. 

To re-establish single electron tunneling, the energy transferred to
the cavity can be regained as displayed in part (iii) of
Fig. \ref{fig1}(b) where the cavity phonon is reabsorbed exciting a
higher lying electronic state
$\mu^*(N+1)$~\cite{footnote:ang_momentum}. To this end, the excited
cavity phonon mode coherently exchanges energy with the electronic
excited states $\mu^*(N+1)$ which can be understood in terms of
(damped) Rabi oscillations~\cite{bib:brandes03}.

An estimate for the phonon modes and energies confined to the quantum
dot cavity 
%(which is orientated in the $\left[ 100 \right]$ direction)
is obtained from microscopic calculations~\cite{bib:debald02} that
find van-Hove singularities in the cavity phonon density of states,
accompanied by an extreme enhancement of phonon emission at certain
phonon energies. The lowest energy where this occurs is $\hbar
\Omega_{\rm ph} \approx 3\hbar c_{\rm L}/z = 73~\mu$eV for quantized
dilatational phonon modes ($\hbar \Omega_{\rm ph} \approx 145~\mu$eV
for flexural modes) for the sample thickness $z = 130$\,nm and the
longitudinal velocity of \hspace{0.1cm} sound \hspace{0.1cm} in the $\hspace{0.1cm} \left[ 100
\right]\hspace{0.1cm}$ direction
of bulk GaAs, $c_{\rm L} = 4.77 \cdot
10^5$\,cm/s. This prediction for $\hbar \Omega_{\rm ph}$, based on a
simple infinite thin-plate model, compares relatively well to the
observed energy gap $\epsilon=100\,\mu$eV found in our transport data
at an electron temperature of $100$\,mK. 
At larger temperatures the electrons gain enough energy to
overcome phonon blockade when the
broadening of the Fermi distribution function in the leads approaches $4k_{\rm
B}T \approx \epsilon_0$. This is observed in Fig. \ref{fig2}(c), where
clear single electron tunneling resonances are found again at $T =
350$\,mK and $B = 0$\,T.
We therefore conclude that quantized cavity
phonons are excited within the cavity.

The phonon gap $\epsilon$ is recorded with a higher resolution in
Fig. \ref{fig3}(a) showing the central region of Fig. \ref{fig2}(a) in
the same color scale. The conventional shape of the Coulomb diamonds
is marked by solid lines. Clearly, the asymmetric shape of the gap can
be discerned: The crossing of the two solid lines indicates the
position of the missing single electron tunneling conductance
peak. The onset of conductance (*) occurs at a slightly different
value of $V_{\rm g}$ by which the half diamonds are offset as
indicated by dotted lines. 
This offset is also explained by the model
developed in Fig. \ref{fig1}(b): At the onset of conductance at
$V_{\rm sd} = \epsilon/ e$ the energy level $\mu(N+1)$ only remains in
the transport window after the level mismatch when initially being
aligned with the upper reservoir. Hence, the initial level position is
shifted by $\delta V_{\rm g} = \epsilon / 2e \alpha$ (where $\alpha =
C_{\rm g} / C_\Sigma$) on the gate voltage axis.

Further evidence for phonon blockade is given by tracing the magnetic
field dependence of the conductance gap. Energetically higher lying
electronic states $\mu^*(N+1)$ with finite angular momentum $\ell \cdot \hbar $
($\ell=1,2,...$) can be brought into resonance with the cavity phonon
by altering the magnetic field such that the excitation energy
$\mu^*(N+1,B) - \mu(N+1,B=0) = \epsilon_0$, corresponding to the
situation displayed in level diagram (iii) of Fig. \ref{fig1}(b). In
this case, the higher states offer a broad cross section for the
emitted phonons to be reabsorbed, allowing for the excitation of the
electron into $\mu^*(N+1)$ thus enabling it to sustain single electron
tunneling. 
Examples for such resonances can be seen at $170$\,mT and
$450$\,mT shown in Fig. \ref{fig3}(b) and (d) whereas a non-resonant
situation at $260$\,mT is displayed in Fig. \ref{fig3}(c). Since
reabsorption of the phonon only occurs with a finite probability, the
height of the conductance peak at zero bias, which is again displayed
in the right part of the figure, is reduced compared to the onset of
conductance through the original ground state at $V_{\rm sd} =
\epsilon / e$ marked by the dotted line as described before.

\begin{figure}[t]
\includegraphics[width=7.5cm]{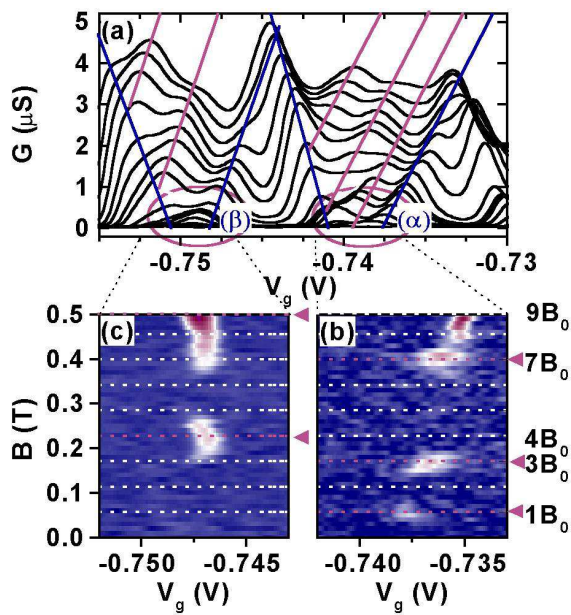}
\caption{
(a) Line plot of conductance resonances $\alpha$ and $\beta$ at
different source-drain bias voltages between $0\,\mu$V and
$-800\,\mu$V. Blue lines follow the ground states, while red lines
mark excited states. 
The conductance at $V_{\rm sd} = 0\,\mu$V is suppressed. (b)
Zero bias conductance for
resonance $\alpha$ plotted against gate voltage $V_{\rm g}$ and
magnetic field $B$. 
Finite conductance
appears for $57$\,mT, $170$\,mT, and $400$\,mT. (c) Similar plot for
resonance $\beta$ (blue:
$0.02\,\mu$S, red: $2\,\mu$S): Non-zero conductance is found for
$230$\,mT and $510$\,mT.
}
\label{fig4}
\vspace{-0.3cm}
\end{figure}

A direct comparison of the magnetic field dependence to transport
spectroscopy on the dot is depicted in Fig. \ref{fig4} where we
consider two adjacent resonance peaks $\alpha$ (right,
c.f. Fig. \ref{fig3}) and $\beta$ (left). The conductance traces are
recorded for bias voltages from $0\,\mu$V to $-800\,\mu$V at $B =
0$\,mT in Fig. \ref{fig4}(a) showing excited states marked by red
lines. The zero bias conductance is plotted logarithmically as a
function of both gate
voltage $V_{\rm g}$ and magnetic field $B$ for $\alpha$ and $\beta$ in
Fig. \ref{fig4}(b) (blue: $0.02\,\mu$S, red: $0.2\,\mu$S) and (c)
(blue: $0.02\,\mu$S, red: $2\,\mu$S), respectively. For the right
resonance $\alpha$ we find excited states (red lines) at energies
$\mu_{\alpha}^{*1} = 230~\mu$eV, $\mu_{\alpha}^{*2} = 440~\mu$eV, and
$\mu_{\alpha}^{*3} =740~\mu$eV. These three excited states correspond
to three magnetic fields permitting zero bias conductance at $57$\,mT,
$170$\,mT, and $400$\,mT. For the left resonance $\beta$ we find
excited states at energies $\mu_{\beta}^{*1} = 380~\mu$eV and
$\mu_{\beta}^{*2} = 760~\mu$eV matching with two magnetic fields
re-enabling zero bias conductance at $230$\,mT and $510$\,mT.  Above
$500$\,mT conductance fully re-emerges while below it is found only at
a set of discrete values as discussed above. These values exactly
correspond to multiples $B = n \cdot B_0$ ($n = 1,3,4,7$ and $9$) of
the first resonance in $\alpha$ observed at $B_0 = 57$\,mT.

The presented measurements demonstrate that for freely suspended
quantum dot cavities single electron tunneling gives rise to the
excitation of a longitudinal cavity phonon. The resulting energy loss
leads to a suppression of linear electron transport and to the
formation of a distinct energy gap. This phonon blockade effect can be
overcome at bias voltages large enough to bridge the energy gap, or at
a sufficiently high bath temperature. A third mechanism circumventing
phonon blockade is given by aligning higher lying electronic states
with distinct angular momentum such that electronic transport is
enabled through these states after reabsorption of the cavity phonon
in a process similar to Rabi oscillations. Finally, placing the dot
into the whole bulk crystal instead of a free-standing cavity
eliminates the blockade effect so that elastic single electron tunneling
is restored.

We like to thank Stefan Debald for stimulating discussions. Support
from the Bundesministerium f\"ur Forschung und Technologie and the
Deutsche Forschungsgemeinschaft is acknowledged.


\begin{thebibliography}{10}
\bibitem{footnote:ang_momentum}
In our case, higher lying electronic
states $\mu^*(N+1)$ with non-zero angular momentum $\ell \cdot \hbar $
($\ell=1,2,...$) are considered the excitation energy of which can be
brought into resonance with the cavity phonon by means of an applied
perpendicular magnetic field.
\end{thebibliography}
\end{document}